\documentclass[a4paper,11pt]{article}
\pdfoutput=1 

\usepackage{jheppub} 

\usepackage[T1]{fontenc} 

\usepackage{color}

\title{\boldmath Features and stability analysis of non-Schwarzschild black hole in quadratic gravity}

\author[a,b,c,*]{Yi-Fu Cai,}
\author[a,c]{Hezi Zhang,}
\author[a,d]{Junyu Liu,}
\author[a,c]{Gong Cheng,}
\author[e,a,*]{Min Wang,}

\affiliation[a]{CAS Key Laboratory for Research in Galaxies and Cosmology, Department of Astronomy, University of Science and Technology of China, Chinese Academy of Sciences, Hefei, Anhui 230026, China}
\affiliation[b]{Department of Physics, McGill University, Montr\'eal, Quebec H3A 2T8, Canada}
\affiliation[c]{School of Physical Sciences, University of Science and Technology of China, Hefei, Anhui 230026, China}
\affiliation[d]{School of the Gifted Young, University of Science and Technology of China, Hefei, Anhui 230026, China}
\affiliation[e]{Faculty of Materials and Energy, Southwest University, Chongqing 400715, China}

\emailAdd{yifucai@ustc.edu.cn}
\emailAdd{zhhz@mail.ustc.edu.cn}
\emailAdd{junyu@mail.ustc.edu.cn}
\emailAdd{cghope@mail.ustc.edu.cn}
\emailAdd{minwang@swu.edu.cn}

\abstract{Black holes are found to exist in gravitational theories with the presence of quadratic curvature terms and behave differently from the Schwarzschild solution. We present an exhaustive analysis for determining the quasinormal modes of a test scalar field propagating in a new class of black hole backgrounds in the case of pure Einstein-Weyl gravity. Our result shows that the field decay of quasinormal modes in such a non-Schwarzschild black hole behaves similarly to the Schwarzschild one, but the decay slope becomes much smoother due to the appearance of the Weyl tensor square in the background theory. We also analyze the frequencies of the quasinormal modes in order to characterize the properties of new back holes, and thus, if these modes can be the source of gravitational waves, the underlying theories may be testable in future gravitational wave experiments. We briefly comment on the issue of quantum (in)stability in this theory at linear order.}

\begin{document}
\maketitle
\flushbottom

\section{Introduction}\label{intro}

Since Einstein's proposal in 1915, General Relativity (GR) has been established as the standard theory of gravitation for one hundred years. As a pillar of modern science, the predictions of GR have been confirmed in all observations to date. However, one of the most challenging task that theoretical physicists are facing today is how GR can be reconciled with the laws of quantum physics to produce a consistent ultraviolet (UV) complete theory of quantum gravity.

To address this issue as well as to be of phenomenological interest, extensions of Einstein gravity with the presence of higher order derivative terms arise in fundamental theories, such as string theory, loop quantum gravity, asymptotically safe gravity and others. In particular, it was found in \cite{Stelle:1976gc} that adding quadratic curvature terms could improve the renormalizability of the underlying gravitational theory although this theory would suffer from an instability of ghost degrees of freedom \cite{Stelle:1977ry}. As shown in \cite{Hawking:2001yt}, however, if the path integral quantization is evaluated in Euclidean space and then Wick rotated to Lorentzian space, this path integral can yield a theory of quantum gravity without a negative norm state. This approach has been applied into inflationary cosmology and provided an interesting interpretation for primordial perturbations \cite{Clunan:2009er}.

Black hole physics is believed to provide an important window to the quantum nature of gravity. Recently, there has been increasing interests in studying black hole solutions in gravity theories by taking into account higher order curvature terms, such as from the perspective of string theory \cite{Boulware:1985wk, Kanti:1995vq, Alexeev:1996vs, Moura:2006pz}, in the Gauss-Bonnet extended gravity \cite{Wheeler:1985nh, Torii:1996yi, Guo:2008hf, Guo:2008eq, Pani:2009wy}, the Einstein-Weyl gravity \cite{Frolov:2009qu, Nelson:2010ig, Lu:2015cqa, Lu:2015psa}, the $f(R)$ gravity \cite{Sebastiani:2010kv, PerezBergliaffa:2011gj, Cognola:2015wqa, Duplessis:2015xva}, and other cases of general quadratic gravity \cite{Jacobson:1993xs, Cvetic:2001bk, Kehagias:2015ata, Alvarez-Gaume:2015rwa, Cognola:2015uva}, as well as the analyses of gravitational energy of quadratic gravity \cite{Boulware:1983td, Boulware:1985nn, Deser:2002rt, Deser:2007vs}. Moreover, it is important to examine the stability issue of a quadratic gravity theory by analyzing linear perturbations, such as in \cite{Myung:2013doa}.

In addition, black hole solutions were obtained in a gravitational theory involving higher order terms within the scenario of asymptotical safety \cite{Cai:2010zh} and its stability issue was addressed in \cite{Liu:2012ee} by analyzing the so-called quasinormal modes. This study provides a representative example to show how the (in)stability issue of a black hole solution could be investigated via the method of analyzing quasinormal modes, of which the generation is due to the quasinormal ringing of the background spacetime under perturbations and hence is associated with the characteristics of black hole. The identification of the quasinormal modes is considered to possibly falsify various black hole solutions derived in a large class of gravity theories through the imminent gravitational wave surveys. Note that the investigation of black hole perturbations has drawn a lot of interest for decades since it is associated with black hole stability, gravitational wave detection as well as some fundamental symmetries such as the gauge/gravity duality. Analyses of these modes in theories of higher derivative gravity were performed in \cite{Iyer:1989rd, Konoplya:2004xx, Abdalla:2005hu, Konoplya:2008ix}. We refer to Refs.~\cite{Berti:2009kk, Konoplya:2011qq} and references therein for recent comprehensive reviews.

In the present paper we aim at examining the quasinormal modes seeded by a test scalar field propagating in black hole solutions of a pure Einstein-Weyl gravity as derived in ref.~\cite{Lu:2015cqa}. In Section \ref{sec:model} we briefly review the background gravitational theory and the associated black hole solutions. Then in Section \ref{sec:quasi} we perform a detailed analysis of quasinormal modes seeded by a massless test scalar field that is propagating freely within various black hole solutions beyond Schwarzschild and investigate their behaviors. Section \ref{sec:quantum} is devoted to a brief discussion of the field equations for gravitational waves, in which one obtains two poles on the dispersion relation and hence this implies an instability of quantum fluctuations. We eventually summarize our results with a discussion in Section \ref{sec:conclusion}. Throughout the whole paper we use geometrized units with $G=c=1$ and the $(-,+,+,+)$ convention for the metric.

\section{Quadratic gravity and black holes beyond Schwarzschild}
\label{sec:model}

We start with a brief introduction to a general theory of quadratic gravity. Consider a four-dimensional Einstein gravity involving quadratic curvature terms of which the action is written as (following ref.~\cite{Lu:2015cqa})
\begin{eqnarray}\label{S_grav}
 {\cal S}_{\rm Grav.} = \int d^4x \frac{\sqrt{-g}}{16\pi} \big( R -\alpha C_{\mu\nu\rho\sigma} C^{\mu\nu\rho\sigma} +\beta R^2 \big)~,
\end{eqnarray}
where
\begin{eqnarray}
 C_{\mu\nu\rho\sigma} \equiv R_{\mu\nu\rho\sigma} - g_{\mu[\rho} R_{\sigma]\nu} + g_{\nu[\rho} R_{\sigma]\mu} + \frac{R}{3}g_{\mu[\rho}g_{\sigma]\nu} ~,
\end{eqnarray}
is defined as the Weyl tensor in four dimension with $R$ being the Ricci scalar. In the above action we have introduced two model parameters $\alpha$ and $\beta$, which are of area dimension: $[L^2]$, and, describe the deviations from the regular Einstein gravity.

\subsection{Spherically symmetric and static solutions}

In order to find a black hole solution beyond Schwarzschild, we consider a static and spherically symmetric ansatz as follows,
\begin{eqnarray}\label{metric_ansatz}
 ds^2 = -N(r)dt^2 +\frac{dr^2}{F(r)} +r^2 d\Omega_2^2 ~,
\end{eqnarray}
in which two dimensionless metric factors $N$ and $F$ have been introduced as functions of the radial coordinate. As was shown in \cite{Nelson:2010ig} as well as argued in \cite{Lu:2015cqa}, the Ricci scalar vanishes for any static black hole solutions of the action \eqref{S_grav}. As a result, the general theory of quadratic gravity can reduce to the pure Einstein-Weyl gravity at the classical level by setting $\beta = 0$.

We refer to the appendix of the present paper for the details of studying a black hole solution beyond Einstein analytically. In the main context, we simply summarize the steps of constructing such a solution as follows. Firstly, one can vary the action with respect to the metric factors and then derive the field equations for $N$ and $F$. Secondly, in order to exhibit the difference between this solution and the Schwarzschild one, one can parameterize the leading term of the $F$ factor (denoted by $F_1$ introduced in \eqref{sol_nhl} in the appendix) as follows,
\begin{equation}\label{F_nhl}
 F_1 = \frac{1+\delta}{r_H}~,
\end{equation}
where $r_H$ represents for the position of the black hole horizon and $\delta$ is the amount of deviation from the Schwarzschild solution since in GR we have $\delta = 0$. Such a parametrization can provide a boundary condition for numerically solving the vacuum structure of the underlying gravity theory, which is the last step to implement in the whole construction. In the following subsection we numerically repeat the result of ref.~\cite{Lu:2015cqa} to demonstrate an existence of a black hole beyond Schwarzschild.

\subsection{Numerical estimates}
\label{subsec:numeric bg}

Note that, as has been observed in \cite{Lu:2015cqa}, for each given $\alpha$, the viable value of $r_H$ is bounded. Specifically, if $r_H$ is too small there is no opportunity to form a non-Schwarzschild black hole; however, if $r_H$ is too large the black hole mass would become negative and hence leads to a quantum instability at high energy scales. For example, by setting $\alpha=0.5$, one numerically derives a bound: $0.876 < r_H < 1.143$. For any selected value of $r_H$ in the above bounded interval, there exists only one value of $\delta$ that allows for a healthy non-Schwarzschild black hole. This phenomenon is related to the fine-tuning of $\Gamma_+=0$ in the weak field limit (see the second part of the appendix). In figure~\ref{fg1} we show an exact example of the numerical construction introduced in the present subsection.

\begin{figure}[tbp]
\centering 
\includegraphics[width=.45\textwidth]{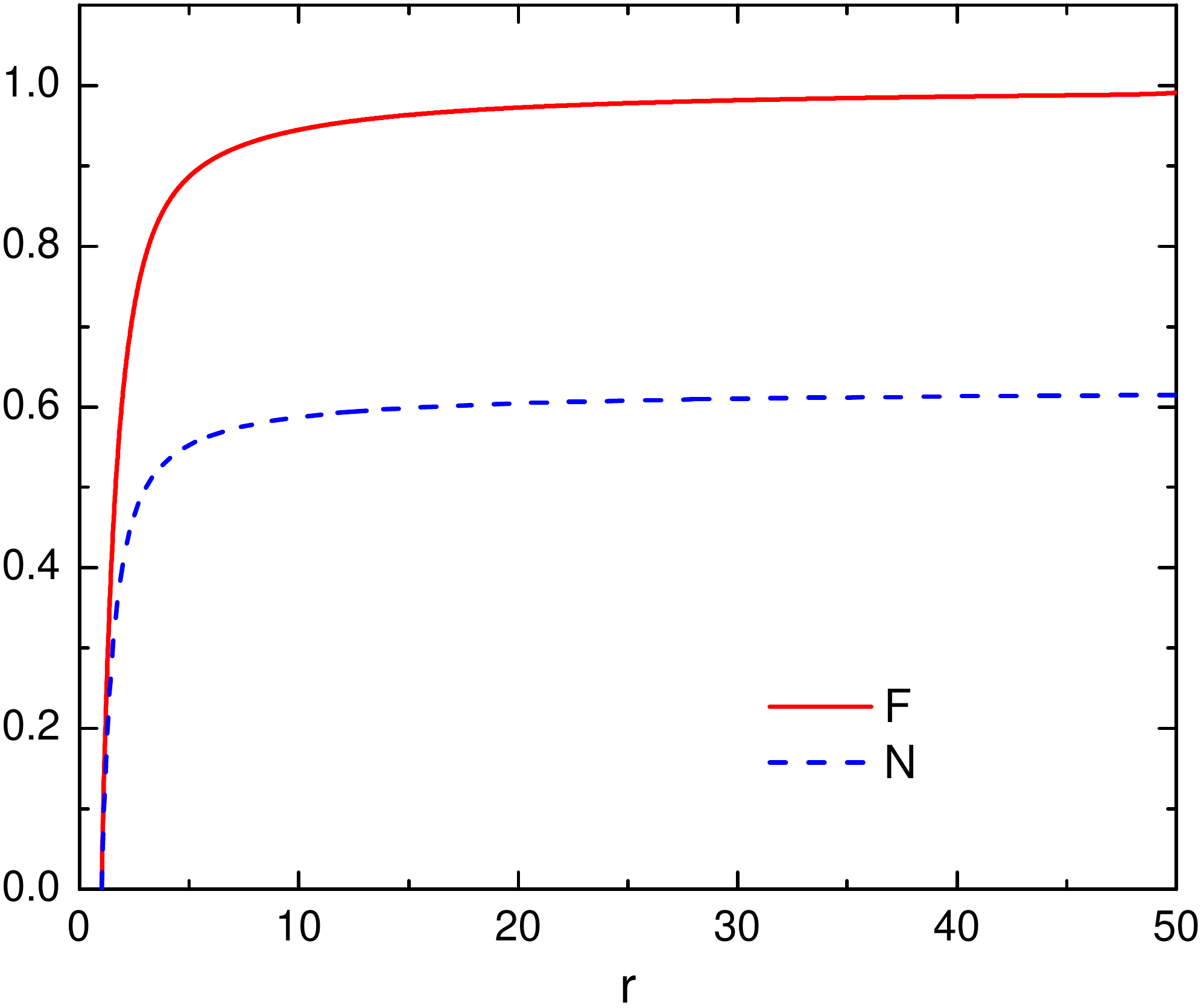}
\caption{\label{fg1} The numerical solution of a non-Schwarzschild black hole. We choose $\alpha=0.5$, $r_H=1.0$, and finely tune $\delta \approx0.3633$. The red, solid curve gives the numerical realization of $F(r)$, while the blue, dashed curve is the numerical result of $N(r)$. Note that, as in the treatment of ref.~\cite{Lu:2015cqa}, we impose the normalization factor to be $0.6$ for $N$ and unity for $F$ at infinity, in order to avoid an asymptotic overlap.}
\end{figure}

\begin{figure}[tbp]
\centering 
\includegraphics[width=.45\textwidth]{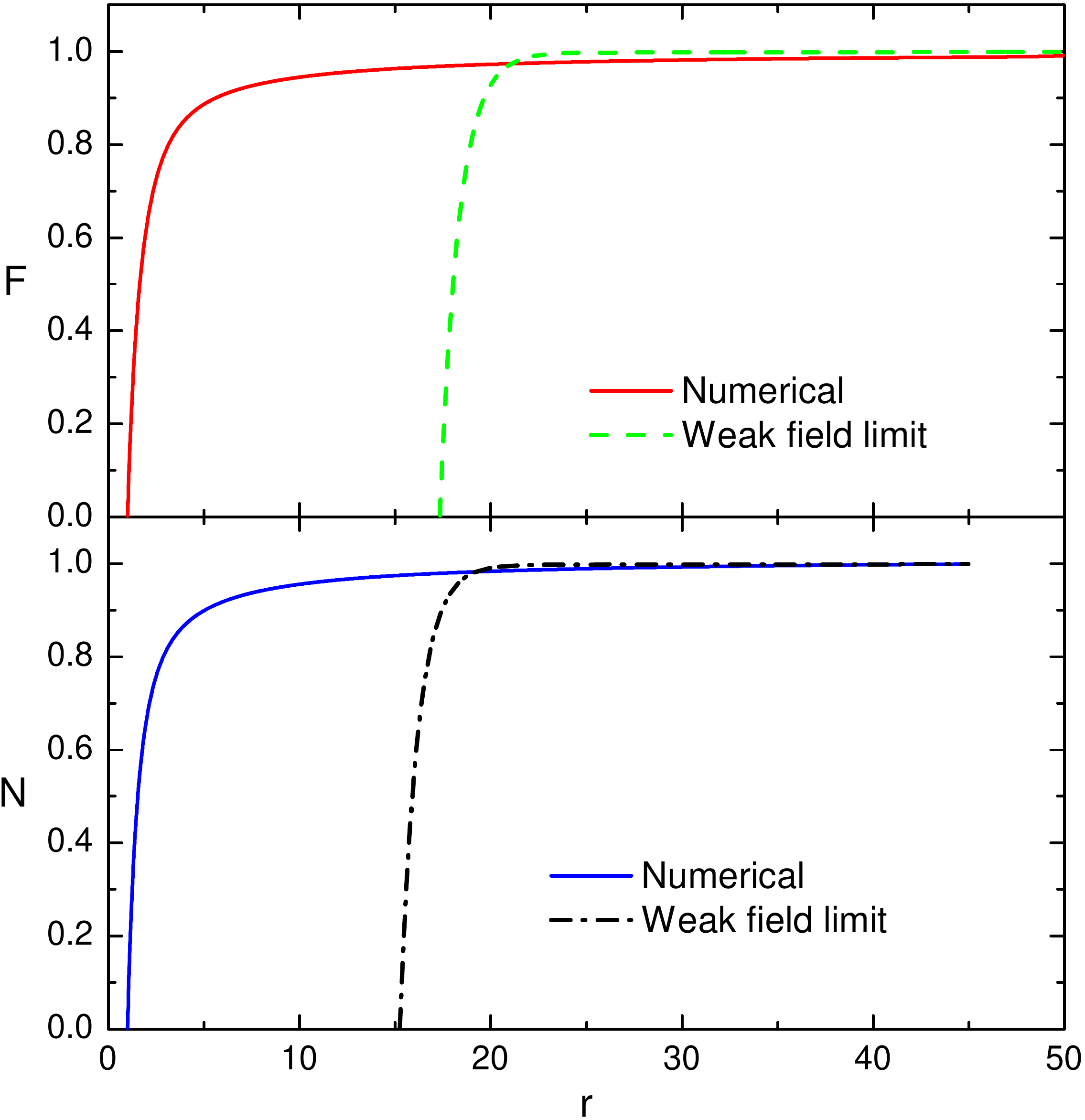}
\caption{\label{fg2} The comparison between the exact solution and the approximate one in \eqref{nf_wfl} under the weak field limit. The model parameters are the same as in figure~\ref{fg1}. The upper panel shows the evolutions of $F$ exactly in numerical computation and approximately in the weak field limit, respectively; and the lower panel describes the dynamics of $N$ in these two cases. }
\end{figure}

Afterwards, using the numerical fitting one can also relate the exact numerical solution to the weak field limit approximate solution in \eqref{sol_wfl} as developed in the appendix. This shows that physically the weak field linearized theory can roughly describe the Einstein-Weyl black hole solution in the large radius regime. A specific fitting is provided in figure~\ref{fg2}. One can read from the figure that the evolution of $F$ and $N$ in the weak field limit is in agreement with the exact numerical results at large radii but deviate badly when evolving to the regime near the horizon.

\section{Quasinormal modes}
\label{sec:quasi}

In this section we study quasinormal perturbations of Einstein-Weyl black holes. Quasinormal modes have been shown to be very useful to uncover intrinsic properties of the geometry (for instance, see \cite{Kokkotas:1999bd}). Consider a massless test scalar field $\psi$ propagating in a black hole background governed by the Einstein-Weyl theory. Its evolution obeys the massless Klein-Gordon equation
\begin{eqnarray}
 \Box\psi = 0~.
\end{eqnarray}

Plugging in the black hole metric into the above Klein-Gordon equation, one gets,
\begin{align}
 \Big( \frac{2F}{r}+\frac{{{\partial }_{r}}F}{2}+\frac{F{{\partial }_{r}}N}{2N} \Big){{\partial }_{r}}\psi
 +F\partial _{r}^{2}\psi -\frac{\partial _{t}^{2}\psi}{N} 
 +\frac{\partial _{\phi }^{2}\psi}{{{r}^{2}}{{\sin }^{2}}\theta } +\frac{{{\partial }_{\theta }}\psi}{{{r}^{2}}\tan \theta } +\frac{\partial _{\theta }^{2}\psi}{{{r}^{2}}} =0~.
\end{align}
In order to solve the above Klein-Gordon equation analytically, it is convenient to use the following standard separation of variables
\begin{eqnarray}
 \psi (t,r,\theta ,\phi ) = \sum_{lm} \frac{1}{r}\Psi_l(r){{Y}_{lm}}(\theta ,\phi ){{e}^{-i\omega t}}~,
\end{eqnarray}
by making use of the spherical harmonic functions. Accordingly, the Klein-Gordon equation can be greatly simplified as:
\begin{align}
 \big( \frac{{{r}^{2}}{{\omega }^{2}}}{N}-\frac{r{{\partial }_{r}}F}{2}-\frac{rF{{\partial }_{r}}N}{2N}-l(l+1) \big) \Psi_l
 +\big( \frac{{{r}^{2}}{{\partial }_{r}}F}{2}+\frac{{{r}^{2}}F{{\partial }_{r}}N}{2N} \big){{\partial }_{r}}\Psi_l+{{r}^{2}}F\partial _{r}^{2}\Psi_l = 0~,
\end{align}
for each fixed value of $l$.

By introducing the generalized tortoise coordinate
\begin{eqnarray}
 d{{r}^{*}}={dr}/{\sqrt{F(r)N(r)}}~,
\end{eqnarray}
we can obtain the Schr\"{o}dinger-type equation for each value of $l$ as follows,
\begin{eqnarray}\label{seq}
 \left( \partial _{r*}^{2}+{{\omega }^{2}}-V_l(r) \right)\Psi_l ({{r}^{*}})=0~,
\end{eqnarray}
where the effective potential is given by
\begin{eqnarray}
V_l(r) = V_l(r(r^*))=\frac{F{{\partial }_{r}}N+N{{\partial }_{r}}F}{2r}+\frac{l(l+1)N}{{{r}^{2}}}~.
\end{eqnarray}
This form can be treated systematically in the analysis of quasinormal modes, which will be given in the following subsections. We will apply the methods of the characteristic integration and the WKB analysis for quantitative estimation of numerical quasinormal modes, respectively, in the following up subsections.

\subsection{Characteristic integration}

A simple but efficient way of solving 1+1 dimensional d'Alembert equations is established in the pioneering work \cite{Gundlach:1993tp}. In this formalism, the standard $(r^*,t)$ coordinates are replaced by the light-cone variables,
\begin{eqnarray}
u=t-r^*~,\,\,\,\,\,\,v=t+r^*~,
\end{eqnarray}
in terms of which all wave equations can have the same form.

Considering the fact that we do not have full analytic solutions for the time-evolving wave equation, one efficient approach is to discretize the mode function as
\begin{align}
 \Phi_l(N) = \Phi_l(W) +\Phi_l(E) -\Phi_l(S) 
  -\frac{{{h}^{2}}}{8}V(S) \big( \Phi_l(W) +\Phi_l(E) \big) +\mathcal{O}({{h}^{4}})~,
\end{align}
where $S=(u,v)$, $W=(u+h,v)$, $E=(u,v+h)$, $N=(u+h,v+h)$, $\Phi=\Psi e^{-i\omega t}$, and $h$ is the discrete step size. Note that, in order to solve the mode function for a fixed $l$, one needs to impose the initial condition at the null boundary $u=u_0$ and $v=v_0$. As will be  confirmed by the following numerical simulations, however, the characteristics of the associated field decay are basically insensitive to the initial conditions imposed.

From now on we would like to drop the subscript $l$ from the mode function $\Phi$ but specify its value in detailed calculations. Through the difference equation mentioned above, one can get a series of time domain data $\Phi(t_0)$, $\Phi(t_0+h)$, $\Phi(t_0+2h)$, etc, for all possible fixed $r^*$, and the quasinormal vibrations can be read-off through the transformation from the time domain to the frequency domain.

\begin{figure}[tbp]
\centering 
\includegraphics[width=.45\textwidth]{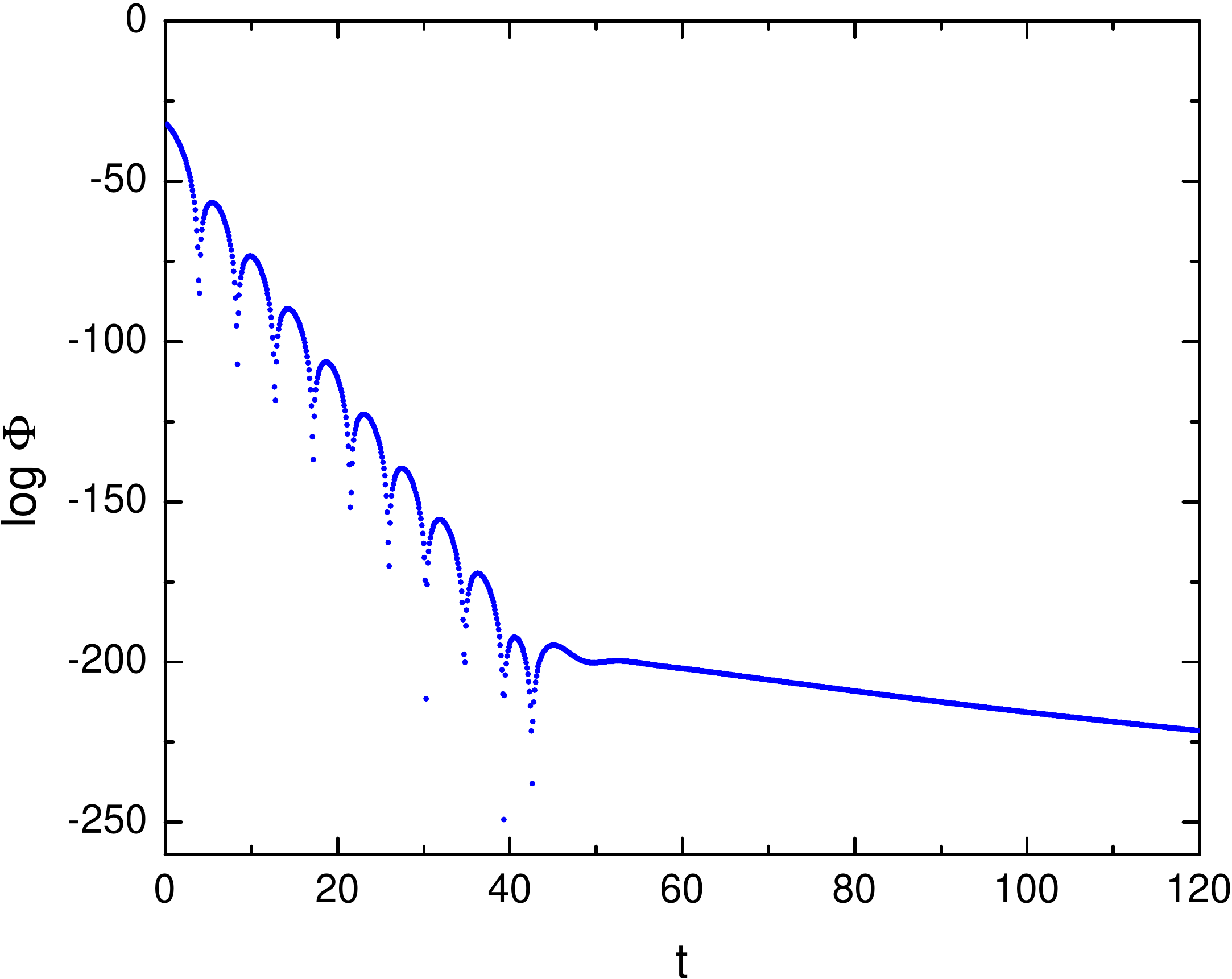}
\caption{\label{fg3} Field decay of the mode function $\log|\Phi(t)|$ for a non-Schwarzschild black hole. We choose $\alpha=0.7$, $r_H=1.1$, and $l=1$ in numerical computations.}
\end{figure}

The numerical computation of the evolution for a mode function $|\Phi|$ is shown in figure~\ref{fg3} on a logarithmic scale. In this plot one can see a representative field decay evolution in the time domain. We choose $\alpha=0.7$, which ensures that the conformal term contribution is smaller than the minimal coupling in the action, and $r_H=1.1$, where exists a numerical non-Schwarzschild solution far from the negative mass region. One can see that the scalar field evolution is firstly dominated by the quasinormal vibration, and then decays with a power-law tail. This is a standard scenario in the time domain profile of black holes in analogue with the case of the Schwarzschild solution (e.g. see the review \cite{Konoplya:2011qq}).

\begin{figure}[tbp]
\centering 
\includegraphics[width=.45\textwidth]{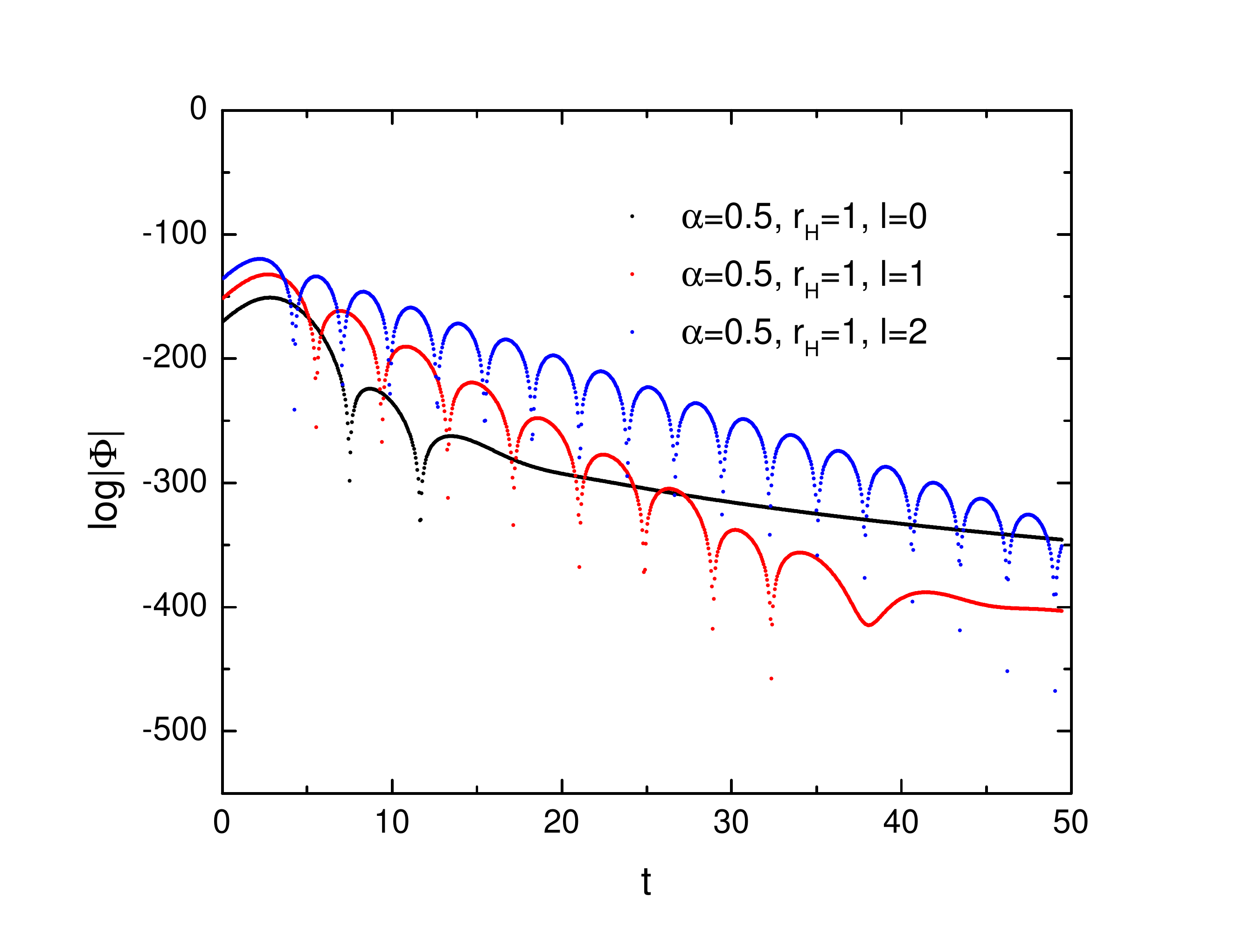}
\includegraphics[width=.45\textwidth]{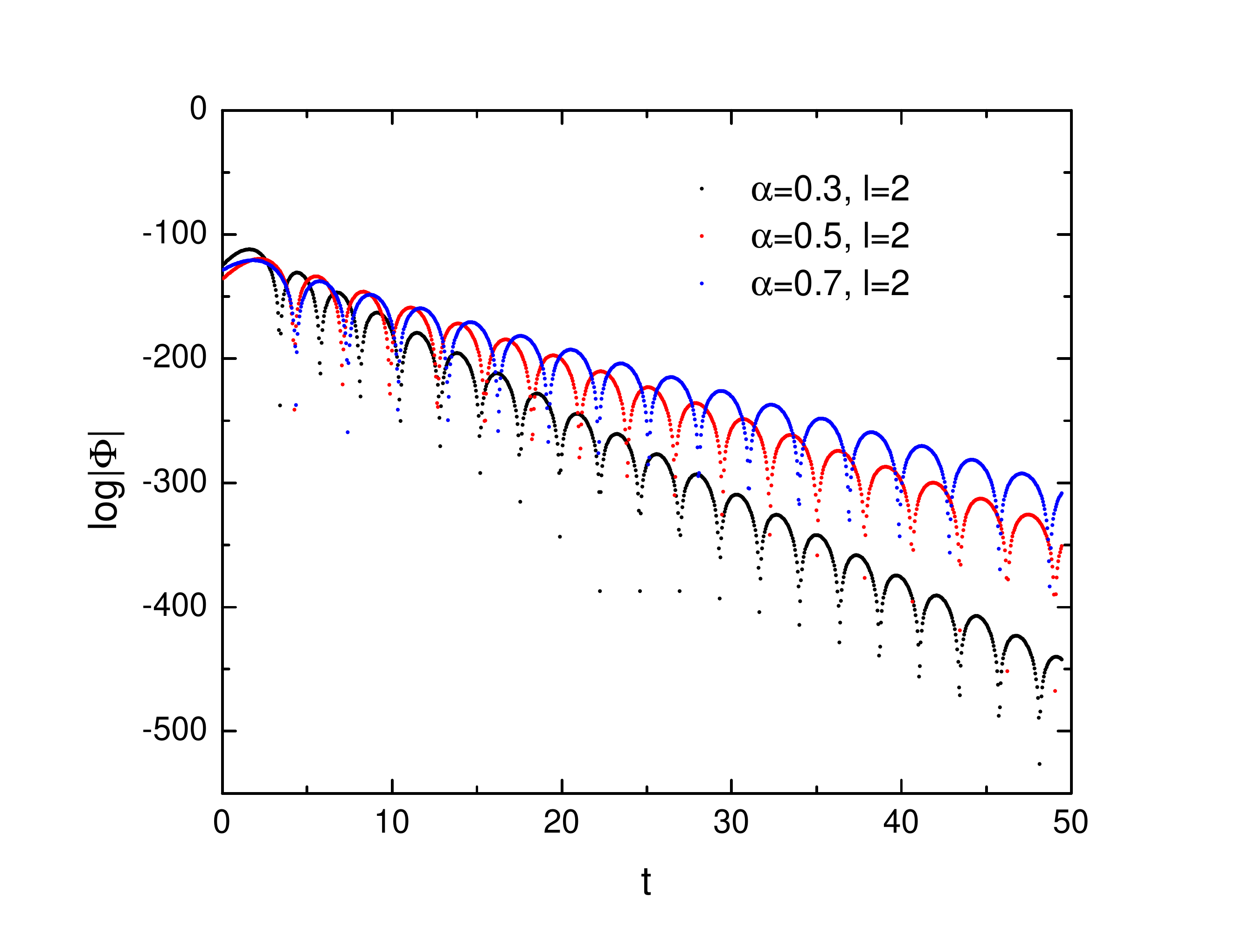}
\caption{\label{fg4}
{
Field decay of $\log|\Phi(t)|$ for different background parameters. In the left panel, we choose $\alpha=0.5$, $r_H=1.0$, with different multipole numbers $l=0,~1,~2$, respectively. In the right panel, we fix $l=2$ and choose different values for $\alpha$, which are: $\alpha=0.3$ ($r_H=0.75$), $\alpha=0.5$ ($r_H=1.0$), $\alpha=0.7$ ($r_H=1.1$). Some power law tails exist in the larger time region, but they are not shown due to the figure size.
}
}
\end{figure}

One can also investigate the field decays with different background parameters as shown in figure~\ref{fg4}. The left panel of figure~\ref{fg4} shows the field decays by varying the value of the multipole number from $0$ to $2$. In this panel, one can read that for larger values of the multipole number $l$, the longer the power-law tail and the larger the slope can be in a given time region. This behavior is also similar to the Schwarzschild case. However, from the right panel of the figure, one can explicitly read that the slope of the field decay strongly depends on the newly introduced model parameter $\alpha$. In the limit $\alpha\rightarrow 0$, it is expected that one can recover the field decay of a Schwarzschild black hole. Moreover, a larger value of $\alpha$ leads to a smoother slope for the field decay for a fixed value of $l$ as shown in the lower panel of figure~\ref{fg4}.

We recall that there exists a bound for the model parameter $\alpha$ that allows for a non-Schwarzschild black hole solution as has been addressed in section \ref{subsec:numeric bg}. In the theory of Einstein-Weyl gravity, this bound is stronger for a smaller value of $\alpha$. Moreover, this bound interval would move along the negative direction of the real axis with $\alpha$ decreasing, so we cannot adjust $r_H$ as a fixed parameter. However, we find that the influence of changing $r_H$ (in the regular bound interval) on the properties of the field decay is not so dramatic as the variation of $\alpha$. Also, with a smaller value of $\alpha$ the field decay evolution has a larger slope with a longer tail, which is similar to the tendency when decreasing $l$. The reason could be understood as follows. The value of $\alpha$ estimates the deviation from the standard Einstein gravity and thus from a standard Schwarzschild black hole, which is the highest symmetric solution (satisfying $F=N$ and $\delta=0$). On the other hand, the multipole number $l$ is also associated with the symmetry of a dynamical system, which is similar to the standard Hydrogen atom problem in quantum mechanics textbooks. Thus, the effects on the field decay by decreasing $\alpha$ and by decreasing $l$ are similar from this perspective.

\subsection{WKB analysis}

In order to understand the quasinormal perturbations (semi-)analytically, it is useful to perform the WKB analysis of the mode function. Imposing $s=i\omega$ on the Schr\"{o}dinger-like equation \eqref{seq}, one gets
\begin{eqnarray}
\frac{{{d}^{2}}\Psi_l }{d{{r}^{*2}}}-({{s}^{2}}+V({{r}^{*}}))\Psi_l =0~.
\end{eqnarray}
This is the Laplace transform of the original time-depended wave equation for $\Psi_l$. Under the boundary condition
\begin{eqnarray}
\underset{{{r}^{*}}\to \pm \infty }{\mathop{\lim }}\,\Psi_l {{e}^{s{{r}^{*}}}}=1~,
\end{eqnarray}
which is very common in usual cases, one can get a discrete set of possible values for $s$($\omega$).

The WKB semi-analytic approach \cite{Schutz:1985zz,Iyer:1986np} is a very successful and efficient method to calculate quasinormal frequencies. In the present study we expand the computation up to the third order, and correspondingly, the square of those frequencies can be written as
\begin{eqnarray}\label{omega_n^2}
 \omega _{n}^{2}=({{V}_{0}}+P)-i\xi {(-2V_{0}^{(2)})^{1/2}}(1+Q) ~,
\end{eqnarray}
where
\begin{equation}\label{P_cal}
 P=\frac{1}{8}\frac{V_{0}^{(4)}}{V_{0}^{(2)}}(\frac{1}{4}+{{\xi }^{2}})-\frac{1}{288}{{\left( \frac{V_{0}^{(3)}}{V_{0}^{(2)}} \right)}^{2}}(7+60{{\xi }^{2}}) ~,
\end{equation}
and
\begin{align}\label{Q_cal}
 -2V_{0}^{(2)}Q = \frac{5}{6912} {{\frac{V_{0}^{(3)4}}{V_{0}^{(2)4}}}} (77 +188{{\xi}^{2}}) 
  -\frac{1}{384} \frac{V_{0}^{(3)2}V_{0}^{(4)}}{V_{0}^{(2)3}} (51 +100{{\xi}^{2}}) +\frac{1}{2304} {{\frac{V_{0}^{(4)2}}{V_{0}^{(2)2}}}} (67+68{{\xi }^{2}}) \nonumber\\
  +\frac{1}{288} \frac{V_{0}^{(3)}V_{0}^{(5)}}{V_{0}^{(2)2}} (19 +28{{\xi}^{2}}) -\frac{1}{288} \frac{V_{0}^{(6)}}{V_{0}^{(2)}} (5 +4{{\xi }^{2}}) ~,
\end{align}
with $\xi=n+{1}/{2}$. In addition, the superscript $(i)$ denotes the $i$-th order differentiation with respect to $r^*$ of the potential $V(r(r^*))$. The subscript 0 means that the potential and its derivatives are calculated at the point $r^*_0$, where $V(r(r^*))$ is an extremum. The solution of $\omega$ can be determined when
\begin{align}
 &n=0,1,2\ldots \,\,\,\,\,\,\,\,\,\,\,\, ~~~~~~~ \text{if }\operatorname{Re}({{\omega }_{n}})>0~,\nonumber\\
 &n=-1,-2,-3\ldots \,\,\,\,\,\,\,\,\,\,\,\,\text{if }\operatorname{Re}({{\omega }_{n}})<0~.
\end{align}
Thus we can use the formula \eqref{omega_n^2}-\eqref{Q_cal} to find the quasinormal modes of the non-Schwarzschild black holes.

\begin{table}[tbp]
\centering
\begin{tabular}{ccc|cc|cc}
\hline\hline
$\alpha$ & $r_H$ & $l$ & WKB:Re($\omega$) & WKB:Im($\omega$) &Int.:Re($\omega$)& Int.:Im($\omega$)
\\
\hline
0.1 & 0.40 & 0 & 0.96413 & -0.98076 & 1.322 & -0.784 \\

0.1 & 0.40 & 1 & 1.57122 & -0.71117 & 1.329 & -0.785 \\

0.1 & 0.40 & 2 & 2.50942 & -0.57118 & 1.901 & -0.532 \\

0.2 & 0.60 & 0 & 1.25506 & -1.36612 & 0.943 & -1.003 \\

0.2 & 0.60 & 1 & 1.14856 & -0.58045 & 1.072 & -0.567 \\

0.2 & 0.60 & 2 & 1.81116 & -0.41710 & 1.505 & -0.372 \\

0.3 & 0.75 & 0 & 0.70972 & -0.76229 & 0.848 & -0.911 \\

0.3 & 0.75 & 1 & 0.94765&  -0.49453 & 0.941 & -0.521\\

0.3 & 0.75 &  2 & 1.49154& -0.37587 & 1.336 & -0.343\\

0.3 & 0.80 & 0 & 0.83671 &  -0.91731 & 0.865 & -0.938\\

0.3 & 0.80 &  1 & 0.97668 &  -0.55954 & 0.950 & -0.543\\

0.3 & 0.80 &  2 & 1.51885  &  -0.42566 & 1.309 & -0.406\\

0.4 & 0.85 & 0 & 0.77774 & -0.84128 & 0.640 &-0.543\\

0.4 & 0.85 & 1 & 0.77737 & -0.30450 & 0.870 &-0.343\\

0.4 & 0.85 & 2 & 1.27841 & -0.26186 & 1.209 &-0.199\\

0.5 & 1.00 & 0 & 0.54346 & -0.58875 & 0.628 & -0.568\\

0.5 & 1.00 & 1 & 0.70183 & -0.29062 & 0.810 & -0.369\\

0.5 & 1.00 & 2 & 1.16591 & -0.31743 & 1.123 & -0.227\\

0.5 & 1.10 &  0 & 0.70962 &  -0.78766 & 0.705 & -0.835\\

0.5 & 1.10 &  1 & 0.81474 &  -0.55477 & 0.782& -0.559\\

0.5 & 1.10 &  2 & 1.19632 &  -0.37069 & 1.112& -0.399\\

0.6 & 1.05 &  0 & 0.50857 &  -0.54633 & 0.562& -0.564\\

0.6 & 1.05 &  1 & 0.65280 &  -0.30540 & 0.768& -0.348\\

0.6 & 1.05 &  2 & 1.05104 &  -0.25422 & 1.097& -0.221\\

0.7 & 1.10 &  0 & 0.37600 &  -0.39482 & 0.504& -0.495\\

0.7 & 1.10 &  1 & 0.59613 &  -0.26813 & 0.739& -0.295\\

0.7 & 1.10 &  2 & 0.96098 &  -0.22188 & 0.710& -0.185\\

0.7 & 1.20 &  0 & 0.47102 &  -0.51351 & 0.558& -0.501\\

0.7 & 1.20 &  1 & 0.63437 &  -0.36070 & 0.732& -0.333\\

0.7 & 1.20 &  2 & 0.97969 &  -0.21523 & 0.985& -0.237\\

0.8 & 1.30 &  0 & 0.41946 &  -0.45870 & 0.539& -0.462\\

0.8 & 1.30 &  1 & 0.54557 &  -0.19984 & 0.703& -0.310\\

0.8 & 1.30 &  2 & 0.92753 &  -0.25469 & 0.971& -0.187\\

1.0 & 1.50 &  0 & 0.29276 &  -0.32336 & 0.508& -0.501\\

1.0 & 1.50 &  1 & 0.54461 &  -0.32918 & 0.653& -0.344\\

1.0 & 1.50 &  2 & 0.83938 &  -0.24689 & 0.915& -0.225\\
\hline\hline
\end{tabular}
\caption{\label{table_QNM}
{
Values for the quasinormal frequencies for the mode function propagating in the non-Schwarzschild geometry based on the WKB method and the algorithm of characteristic integration, respectively. The value of model parameter $\alpha$ varies from $0.1$ to $1.0$ smoothly. The integer value of the multipole number varies from $l=0$ to $2$, respectively.
}
}
\end{table}

The accuracy of the WKB method is sensitive to specific black hole solutions \cite{Konoplya:2003dd}. However, as has been observed in \cite{Abdalla:2005hu, Cardoso:2004fi}, for low overtones ($l>n$) the accuracy becomes better.
In order to compare explicitly the method of the characteristic integration and the WKB analysis, we perform detailed analyses of two methods by varying the value of $\alpha$ from 0.1 to $1.0$ and the multipole number from $l=0$ to $2$, respectively. Our results are presented in table~\ref{table_QNM}.

From this table, we find that the results of the WKB analysis and the characteristic integration results are fairly consistent but not with high accuracy. Regarding this issue, we argue that there exists a limit on the accuracy of both two methods. For the method of the characteristic integration, the error by solving the differential equations could lead the wave function to deviate from the original form of the differential equations and the corresponding boundary conditions could be affected as well. To overcome this numerical deviation, one needs to improve the method of computer algorithm, which is a very detailed technical issue. Moreover, the accuracy of the results obtained from the WKB approach mainly rely on to which order one truncates the computation. Consider that the WKB calculations at higher order would become extremely lengthy and do not change the results qualitatively, we would like to simply take the third order truncation in our detailed analysis.

Some generic features can be concluded here. First, in the regular bounded interval of $r_H$, all quasinormal mode frequencies have a negative imaginary part, which shows that scalar perturbations in the non-Schwarzschild black hole backgrounds is stable in the time evolution\footnote{We only study the parameter region of physical interest, namely, we have required $\alpha\leq 1$ to ensure that the higher order derivative terms are perturbative to the Einstein-Hilbert sector, as well as a positively definite black hole mass.}. Second, it is observed that with increasing $\alpha$, $l$ and decreasing $r_H$, the absolute value of the imaginary part of the frequency decreases. This observation is consistent with our previous argument on the tendency of the slope because the imaginary part of the frequency stands for the slope of the logarithmic time domain decay. Third, the real part of the frequency stands for the trigonometric vibration of the scalar field. This real part dramatically increases along with a larger value of the multipole number $l$.

\section{Field equations for gravitational waves}\label{sec:quantum}

In this section we briefly discuss tensor perturbations around a non-Schwarzschild black hole solution. Here we use Greek letters $\mu, \nu,...$ to denote the coordinates on the four-dimensional spacetime, Latin letters $i,j,k...$ to denote the coordinates on the two-dimensional space submanifold $S^2$ (namely, $(\theta,\phi)$ coordinates), and $r$, $t$ to denote the radial and time coordinate respectively. And also, we use a comma to denote an ordinary derivative, while a semicolon denotes a covariant derivative.

Let us consider linear perturbations about the background metric,
\begin{eqnarray}
{{g}_{\mu \nu }}\to {{g}_{\mu \nu }}+{{h}_{\mu \nu }}~,
\end{eqnarray}
where the transverse and traceless tensor satisfies
\begin{eqnarray}
{{\nabla }^{\mu }}{{h}_{\mu \nu }}=0~,\,\,\,\,\,{{g}^{\mu \nu }}{{h}_{\mu \nu }}=0~.
\end{eqnarray}

The first order perturbation theory will provide simple results for the perturbation of the Riemann tensor, Ricci tensor and Einstein tensor for spherically symmetric and static backgrounds \cite{Dotti:2005sq}. We simply summarize them as follows. By writing the components of the Ricci and Riemann tensors as
\begin{align}
 & {{R}}_{i}^{j}=\mathcal{G}{{\delta }}_{i}^{j}~,~~
 {R}_{ij}^{kl}=\mathcal{M}({{\delta }}_{i}^{k}{{\delta }}_{j}^{l}-{{\delta }}_{j}^{k}{{\delta }}_{i}^{l})~,\nonumber\\
 & R_{ti}^{tj}=\mathcal{T}\delta _{i}^{j}~,~~
 R_{ri}^{rj}=\mathcal{D}\delta _{i}^{j}~,
\end{align}
where
\begin{align}
 \mathcal{G} &=-\frac{N(-2+2F+r{{F}_{,r}})+rF{{N}_{,r}}}{2{{r}^{2}}N}~,\nonumber\\
 \mathcal{M} &=\frac{1-{{F}}}{{{r}^{2}}}~,~
 \mathcal{T} =-\frac{F{{N}_{,r}}}{2rN}~,~
 \mathcal{D} =-\frac{{{F}_{,r}}}{2r}~,
\end{align}
we can derive the perturbations of these geometric tensors
\begin{align}
 \delta {R}_{\mu}^{\nu} &= \frac{1}{2}( -\square h_{i}^{j} +2\mathcal{M}h_{i}^{j})~,\nonumber\\
 \delta R_{t\mu}^{t\nu} &= \frac{1}{2}(-\mathcal{T}h_{i}^{j} -{{\nabla}_{l}}{{\nabla }^{l}}h_{i}^{j} +{{\nabla }_{l}}{{\nabla }^{j}}h_{i}^{l})~, \\
 \delta R_{r\mu}^{r\nu} &= \frac{1}{2}( {{\nabla }_{i}}{{\nabla }^{l}}h_{l}^{j} +{{\nabla}_{l}}{{\nabla }^{j}}h_{i}^{l} -{{\nabla}_{l}}{{\nabla }^{l}}h_{i}^{j} -{{\nabla }_{i}}{{\nabla }^{j}}h_{l}^{l} -\mathcal{D}h_{i}^{j})~. \nonumber
\end{align}
Here the LHS has indexes $(\mu,\nu)$, while the RHS may only have $(i,j)$. This convention means that the related tensors are nontrivial only when $(\mu,\nu)$ are on the submanifold $S^2$. As a result, one gets $\delta R=0$.

After a very lengthy computation, one can derive the field equation for tensor fluctuations at leading order, which is given by
\begin{align}
 & -\frac{1}{4}{{\square }^{2}}h_{i}^{j} +\frac{1}{8} \big( R_{t}^{t} +R_{r}^{r} -6\mathcal{G} +\frac{1}{3}R +4\mathcal{M} +\frac{1}{\alpha} \big) \square h_{i}^{j} \nonumber\\
 & +\frac{1}{4} \Big\{ \Big[ 6\mathcal{G} -\frac{1}{3}R -\frac{1}{\alpha} -(R_{t}^{t}+R_{r}^{r}) \Big]\mathcal{M} +2\Box\mathcal{M} \Big\} h_{i}^{j}\text{ } \nonumber\\
 & +(\frac{1}{2}{{P}^{t\mu }}_{;\mu } +{\mathcal{M}^{;t}}){{\partial }_{t}}h_{i}^{j} +\Big[ \frac{1}{2}{{P}^{r\mu }}_{;\mu }+{\mathcal{M}^{;r}}-\frac{F}{r}(\mathcal{G}-R_{r}^{r}) \Big] {{\partial }_{r}}h_{i}^{j} 
  =0~,
\end{align}
where we have introduced the Schouten tensor
\begin{eqnarray}
 {{P}_{\mu\nu}} = \frac{1}{2}({{R}_{\mu\nu}} -\frac{1}{6}R{{g}_{\mu\nu}})~.
\end{eqnarray}

It is interesting to notice that, there exist the $\square^2$ operator which appears in the first term of the above field equation and cannot be canceled by any constraint equations. This implies that at high energy the tensor fluctuations would obtain an extra degree of freedom. Such a new degree of freedom often corresponds to a ghost mode that would spoil the stability of the vacuum state quantum mechanically, such as in the Lee-Wick theory of particle physics.

In this regard, the Einstein-Weyl theory of gravitation may still suffer from the quantum instability issue even though this theory is classically stable since the classical scalar perturbations can well behave as analyzed in the previous section. This instability was recently also addressed in ref.~\cite{Hinterbichler:2015soa}, where the authors applied the St\"uckelberg approach to show that the interplay between the ghost graviton and the healthy graviton allows the theory to evade the usual strong coupling issue widely existing in massive gravity theories and become renormalizable, at the expense of stability.

\section{Conclusion}
\label{sec:conclusion}

Recently, the theory of quadratic gravity has drawn the interest of theoretical physicists in the literature from various aspects \cite{Conroy:2014eja, Capozziello:2015nga, Maggiore:2015rma, Mauro:2015waa}. In particular, it was found in \cite{Lu:2015cqa} that the Einstein-Weyl gravity can allow for a black hole solution beyond Schwarzschild. In the present work we have revisited this type of new black hole solutions at both the background and perturbative levels. At the background level, we analyzed the solutions of the metric factors in the weak field limit and compared them with the exact numerical results. At the perturbative level, we have studied in detail the propagation of quasinormal perturbations seeded by a test scalar field within such a black hole background.

Specifically, we have analyzed the frequencies and time domain evolutions of quasinormal modes seeded by this massless scalar field in the exterior of such a non-Schwarzschild black hole as derived in the Einstein-Weyl theory of gravity. Our results show that the time domain evolution of the quasinormal modes is similar to that obtained in the regular Schwarzschild case where the mode functions decay in a power-law form. However, due to the existence of the higher derivative term, the slope of the field decay is generally smoother than that in the Schwarzschild one.
In addition, we present a brief analysis of tensor perturbations, which are regarded as gravitational waves. We show explicitly that the linearized field equation for gravitational waves involves higher derivative operators that would bring the theory to be unstable quantum mechanically at high energy scales. However, it would be interesting to study in more depth this issue under non-perturbative approaches in order to reveal the relation between a ghost graviton mode and quantum renormalizability of gravity theories.

\appendix

\section{Appendix}

In the first part of this Appendix, we provide a detailed instruction to the background theory of quadratic gravity. In the second part, we show how a black hole solution beyond Schwarzschild can be obtained in this theory.

\subsection{The theory of quadratic gravity}

Varying the action \eqref{S_grav} with respect to the metric yields the background field equation as follows,
\begin{eqnarray}\label{bg_eom}
 R_{\mu\nu} -\frac{R}{2}g_{\mu\nu} -4\alpha B_{\mu\nu} +2\beta {\cal C}_{\mu\nu}  =0 ~,
\end{eqnarray}
where we have introduced
\begin{eqnarray}
 {\cal C}_{\mu\nu} \equiv R(R_{\mu\nu}-\frac{R}{4}g_{\mu\nu}) +g_{\mu\nu}\Box R-\nabla_\mu\nabla_\nu R
\end{eqnarray}
which is from the conformal gravity part, and also, the Bach tensor
\begin{eqnarray}\label{Bach_tensor}
 B_{\mu\nu} \equiv (\nabla^\rho\nabla^\sigma+\frac{1}{2}R^{\rho\sigma}) C_{\mu\rho\nu\sigma}
\end{eqnarray}
due to the Weyl-Eddington term.

In order to study the vacuum structure of a static spacetime satisfying spherical symmetry, we assume the absence of matter fields in the above system. As was shown in \cite{Nelson:2010ig} as well as argued in \cite{Lu:2015cqa}, the Ricci scalar vanishes for any static black hole solutions of the action \eqref{S_grav}. That is, $R=0$ in the above classical theory. It turns out that, at the classical level, one can greatly simplify the quadratic gravity action \eqref{S_grav} by taking $\beta=0$, and therefore, the generic action reduces to a theory of pure Einstein-Weyl gravity.

As was pointed out in \cite{Lu:2015cqa}, however, the requirement of $R=0$ does not simply lead to the trivial solution of the Schwarzschild black hole. This is because, by setting $R=0$ and integrating the trace of the field equation \eqref{bg_eom} over the spatial region could yield a nontrivial and non-vanishing Ricci tensor of the four-dimensional spacetime, although the surface term remains zero. This is also the key reason that there might exist static and spherically symmetric black holes over and above the Schwarzschild one as shown in \cite{Lu:2015cqa}.

\subsection{Black holes beyond Schwarzschild}

Plugging the spherically symmetric and static ansatz \eqref{metric_ansatz} into the background field equation \eqref{bg_eom} leads to the following two second order differential equations:
\begin{align}\label{eom_FN}
 F_{,rr} =& \frac{1}{(rN_{,r}-2N)} \Big[ - \frac{3F_{,r}^2}{2F}N -F_{,r}N_{,r} -\frac{(3F+rF_{,r})}{2N}N_{,r}^2 \nonumber\\
 & + \frac{rF}{2N^2}N_{,r}^3 + \frac{2N}{r}(1-F) \big(\frac{2}{r}+\frac{F_{,r}}{F}\big) 
  - \frac{1}{\alpha}\big(\frac{N}{F}-N-rN_{,r}\big) \Big]  ~, \\
 N_{,rr} =& \big[\frac{2(1-F)}{r^2F}-\frac{2F_{,r}}{rF}\big]N -\big(\frac{2}{r}+\frac{F_{,r}}{2F}\big)N_{,r} +\frac{1}{2h}N_{,r}^2 ~,
\end{align}
where the subscript $_{,r}$ represents a derivative with respect to $r$. To fully determine the solutions of these two differential equations, one also needs to impose the horizon condition that
\begin{eqnarray}\label{horizon_boundary}
 F(r=r_H) = N(r=r_H) = 0~,
\end{eqnarray}
with $r_H$ being defined as the position of the black hole horizon.

\subsubsection{Near horizon limit}\label{NHL}

Since the above two metric factors vanish at the horizon, in order to grasp the physics of the black hole solution near the horizon, it is convenient to make Taylor expansions as follows,
\begin{eqnarray}\label{sol_nhl}
 F(r)= \sum_{i=1}^{\infty} F_i (r-r_H)^i~, ~~ N(r)= \sum_{i=1}^{\infty} N_i (r-r_H)^i~.
\end{eqnarray}
Note that, among all the coefficients of the Taylor expansions, there exists at least one parameter that is a normalization factor and accordingly can be absorbed by a time re-scaling. In the following we take $N_1$ to be the normalization factor without loss of generality. One can plug these expansions into the field equations \eqref{eom_FN} and solve for $F_i$ and $N_i$ order by order. In addition, in the parametrization \eqref{F_nhl} we have introduced $\delta$ to characterize the difference between this solution and the Schwarzschild one.

\subsubsection{The linearized treatment in weak field limit}\label{WFL}

From the other side, it is well known that the metric factors should approach unity when far away from the black hole in order to be consistent with the boundary condition of Minkowski spacetime. Therefore, one can analyze these metric factors in the weak field limit at large scales. This method has been widely applied in the literature and turned out to be very useful in analyzing black hole systems in modified gravity theories, for instance, in massive gravity models \cite{Koyama:2011yg, Cai:2012db}. In this limit, we can expand the metric factors around a Minkowski background as
\begin{eqnarray}
 F(r) = 1+f(r)~,~~ N(r) = 1+n(r) ~,
\end{eqnarray}
in the limit where $r$ is much larger than the horizon.

Keeping leading order terms in $n$ and $f$, the field equations can then be greatly simplified, of which the forms are given by,
\begin{align}
 r^2f(r) -4\alpha f(r) +r^3n_{,r} +2r^2\alpha f_{,rr} &= 0~, \nonumber\\
 2f(r) +2r f_{,r} +2r n_{,r} +r^2 n_{,rr} &= 0~.
\end{align}
Consequently, the metric factors in the weak field limit can be approximately solved as
\begin{align}\label{sol_wfl}
 n(r) &= \frac{\Gamma_0}{r} +\Gamma_-\frac{e^{-mr}}{r} +\Gamma_+\frac{e^{+mr}}{r} ~, \\
 f(r) &= \frac{\Gamma_0}{r} +\Gamma_-\frac{(1+mr)e^{-mr}}{2r} +\Gamma_+\frac{(1-mr)e^{+mr}}{2r} ~, \nonumber
\end{align}
with $m^2 \equiv 1/2\alpha$ being introduced. Moreover, the coefficients $\Gamma_0$, $\Gamma_{\pm}$ are integral constants that can be determined by numerically evolving the metric factors from the horizon to large length scales.

From the approximate solution in the weak field limit, one can immediately notice that the appearance of the $e^{mr}/r$ term would severely spoil the classical stability of the black hole system governed by the Einstein-Weyl theory. One possible way of circumventing this issue is to finely tune the value of $\delta$ introduced in the parametrization \eqref{F_nhl} to ensure $\Gamma_+ = 0$. Under this condition one gets
\begin{align}\label{nf_wfl}
 n(r) &= \frac{\Gamma_0}{r} +\Gamma_-\frac{e^{-mr}}{r} ~, \nonumber\\
 f(r) &= \frac{\Gamma_0}{r} +\Gamma_-\frac{(1+mr)e^{-mr}}{2r} ~.
\end{align}
This limit also shows that we cannot choose a negative $\alpha$ for a regular black hole solution, in which case $m$ will be imaginary and cause a vibration at large radii for $f(r)$. Since for a specific numerical solution as will be constructed in the next section one can numerically fits the values of $\Gamma_0$ and $\Gamma_-$ by matching the linearized solution with the exact one.

\acknowledgments

We are grateful to R. Brandenberger, D. Liu, J. Quintin, Z. Wang, and M. Zhu for valuable comments. We particularly thank H. L\"u for extensive discussions and insightful suggestions during the study of this project. YFC is supported in part by the Chinese National Youth Thousand Talents Program, by the USTC start-up funding (Grant No.~KY2030000049) and by the National Natural Science Foundation of China (Grant No. 11421303). The works of GC, JL and HZ are supported in part by the USTC program of national science talent training base in astrophysics. MW is supported by the National Natural Science Foundation of China (Grant No.~21203154).

\end{document}